\documentclass[pre,aps,superscriptaddress,floatfix]{revtex4}
\usepackage{amsmath}
\usepackage{amssymb}
\usepackage{amsbsy}
\usepackage{graphicx}
\usepackage{color}

\usepackage{enumerate}
\usepackage{mathtools}

\def\be{\begin{equation}}
\def\ee{\end{equation}}
\def\bfi{\begin{figure}}
\def\efi{\end{figure}}
\def\bea{\begin{eqnarray}}
\def\eea{\end{eqnarray}}

\usepackage{environ}
\NewEnviron{myequation1}{%
\begin{equation}
\scalebox{0.95}{$\BODY$}
\end{equation}
}
\NewEnviron{myequation2}{%
\begin{equation}
\scalebox{0.98}{$\BODY$}
\end{equation}
}
\NewEnviron{myequation3}{%
\begin{equation}
\scalebox{0.9}{$\BODY$}
\end{equation}
}
\NewEnviron{myequation4}{%
\begin{equation}
\scalebox{0.86}{$\BODY$}
\end{equation}
}
\NewEnviron{myequation5}{%
\begin{equation}
\scalebox{0.85}{$\BODY$}
\end{equation}
}
\NewEnviron{myequation6}{%
\begin{equation}
\scalebox{0.9}{$\BODY$}
\end{equation}
}

\DeclareMathOperator\erf{erf}
\DeclareMathOperator\erfc{erfc}

\newcommand{\der}{\frac{d}{dt}}

\newcommand{\infinito}{\mathcal{1}}

\newcommand{\TR}{T_{  R}}
\newcommand{\TL}{T_{  L}}

\newcommand{\zL}{z_{  L}}
\newcommand{\zR}{z_{  R}}

\newcommand{\MB}{\sqrt{\frac{m}{2\pi k_{B} T}}e^{-\frac{m}{2k_{  B} T}v^2}}

\newcommand{\MBL}{\sqrt{\frac{m}{2\pi k_{\scriptscriptstyle B} \TL}}e^{-\frac{m}{2k_{\scriptscriptstyle B} \TL}v^2}}
\newcommand{\MBR}{\sqrt{\frac{m}{2\pi k_{\scriptscriptstyle B} \TR}}e^{-\frac{m}{2k_{\scriptscriptstyle B} \TR}v^2}}

\newcommand{\URTODX}{(v-V)\Theta(v-V)dt}
\newcommand{\URTOSX}{(V-v)\Theta(V-v)dt}

\newcommand{\intpiu}{\int^{\mathcal{1}}_{V}}
\newcommand{\intmeno}{\int_{-\mathcal{1}}^{V}}
\newcommand{\intpiumeno}{\int_{-\mathcal{1}}^{\mathcal{1}}}

\newcommand{\norm}{\frac{m}{2\,k_{B} T}}
\newcommand{\normL}{\frac{m}{2\,k_{B} \TL}}
\newcommand{\normR}{\frac{m}{2\,k_{B} \TR}}

\newcommand{\invnorm}{\frac{2\,k_{\scriptscriptstyle B} T}{m}}
\newcommand{\invnormL}{\frac{2\,k_{\scriptscriptstyle B} \TL}{m}}
\newcommand{\invnormR}{\frac{2\,k_{\scriptscriptstyle B} \TR}{m}}

\newcommand{\factVqL}{\frac{MN}{\Delta \zL \sqrt{\pi}} \frac{4m}{m+M}}
\newcommand{\factVqR}{\frac{MN}{\Delta \zR \sqrt{\pi}} \frac{4m}{m+M}}

\newcommand{\factVq}{MN \frac{4m}{m+M}}

\newcommand{\intpiuR}{\int^{\mathcal{1}}_{V_R}}
\newcommand{\intpiuL}{\int_{-\mathcal{1}}^{V_L}}

\newcommand{\KB}{k_{  B}}

\newcommand{\TTL}{T_{A}}
\newcommand{\TTR}{T_{B}}

\begin{document}

\title{Fourier's Law in a Generalized Piston Model}

\author{Lorenzo Caprini}
\affiliation{GSSI (Gran Sasso Science Institute), Universit\`a dell'Aquila, Viale Francesco Crispi, 7, 67100 L'Aquila, Italy.}

\author{Luca Cerino}
\affiliation{Dipartimento di Fisica, Universit\`a di Roma Sapienza, P.le Aldo Moro 2, 00185 Rome, Italy.}

\author{Alessandro Sarracino}
\affiliation{ISC-CNR and Dipartimento di Fisica, Universit\`a di Roma Sapienza, P.le Aldo Moro 2, 00185, Rome, Italy.}

\author{Angelo Vulpiani}
\affiliation{Dipartimento di Fisica, Universit\`a di Roma Sapienza, P.le Aldo Moro 2, 00185 Rome, Italy.}
\affiliation{Centro Interdisciplinare B. Segre, Accademia dei Lincei, Via della Lungara 10, 00165 Rome, Italy.}

\begin{abstract}
A simplified, but non trivial, mechanical model---gas of $N$ particles
of mass $m$ in a box partitioned by $n$ mobile adiabatic walls of mass
$M$---interacting with two thermal baths at different temperatures, is
discussed in the framework of kinetic theory.  Following an approach
due to Smoluchowski, from an analysis of the collisions
particles/walls, we derive the values of the main thermodynamic
quantities for the stationary non-equilibrium states.  The results are
compared with extensive numerical simulations; in the limit of large
$n$, $mN/M\gg 1$ and $m/M \ll 1$, we find a good approximation of
Fourier's law.
\end{abstract}

\maketitle

\section{Introduction}
\label{zero}

Fourier's law, which relates the macroscopic heat flux to the
temperature gradient in a solid system, was introduced almost two
centuries ago.  Nevertheless, its understanding from microscopic basis
is still an important open issue of the non-equilibrium statistical
mechanics~\cite{LLP03}. In particular, among the several theoretical
studies on this subject, important results have been derived mainly
for $1d$ systems. The prototype model is a chain of masses and non
linear springs (Fermi-Pasta-Ulam-like systems), whose ends interacts
with thermal baths at different
temperatures~\cite{GHN01,D01,LLP03}. Other investigated systems are
constituted of $1d$ lattices~\cite{LLP98,GLPV00}, chains of cells,
with an energy storage device, which exchange energy through tracer
particles~\cite{EY04,EY05}, spinning disks~\cite{SLL14}, 
  systems with local thermalization mechanism~\cite{GG08,GG08b}, or
  a chain of anharmonic oscillators with local energy conserving
  noise~\cite{OS13}. Despite the apparent simplicity of the problem,
and of the considered models, both the analytical approaches and the
numerical studies are rather difficult in this context.

The main aim of the present paper is the study of Fourier's law using
a mechanical model, which is rather crude (but still non-trivial),
allowing for an approach in terms of kinetic theory.  More
specifically, we consider a generalized piston model, made of a
certain number of cells, each containing a non-interacting particle
gas. The walls of the cells are mobile, massive objects, that interact
with the particles via elastic collisions.  The system at its ends
interacts with thermal baths at fixed temperatures.  It is easy to
realize the analogy between such a generalized piston model and the
systems of masses and springs: the pistons and the gas compartments
play the role of masses and springs, respectively.

Our model is an example of partitioning system (as the adiabatic
piston), where previous studies showed that the presence of mobile
walls can induce interesting
behaviours~\cite{CPS96,GP99,GPL03,GL05,BKM04,CPPV07,DCPS11,SGP13,GCSVV,CPV15}.
Basically, in the study of partitioning systems, one can adopt two
approaches: in terms of a Boltzmann equation~\cite{GPL03} or introducing effective
equations (Langevin-like) for suitable observables derived \`a la
Smoluchowski, i.e., from an analysis of the collisions
particles/walls.  In our study, we will adopt the latter method.

The paper is organized as follows. Section 1 is devoted to the
introduction of the model; in Section~2, we show how a thermodynamic
approach is not sufficient to determine the values of macroscopic
variables in the steady state. In Section 3, we present a
coarse-grained Smoluchowski-like description of the system, which
provides a good prediction for the main quantities of interest. \mbox{In
Section 4}, we compare the theoretical results with numerical
simulations and discuss the limit of validity of the proposed
approach. In Section 5, some general conclusions on Fourier's Law
in mechanical systems are drawn. In Appendix A, we report the details
of the analytical computations presented in Section 3.

\section{A Generalized Piston Model}
\label{one}

\begin{figure}[ht!]
\centering
\includegraphics[width=.5\columnwidth,clip=true]{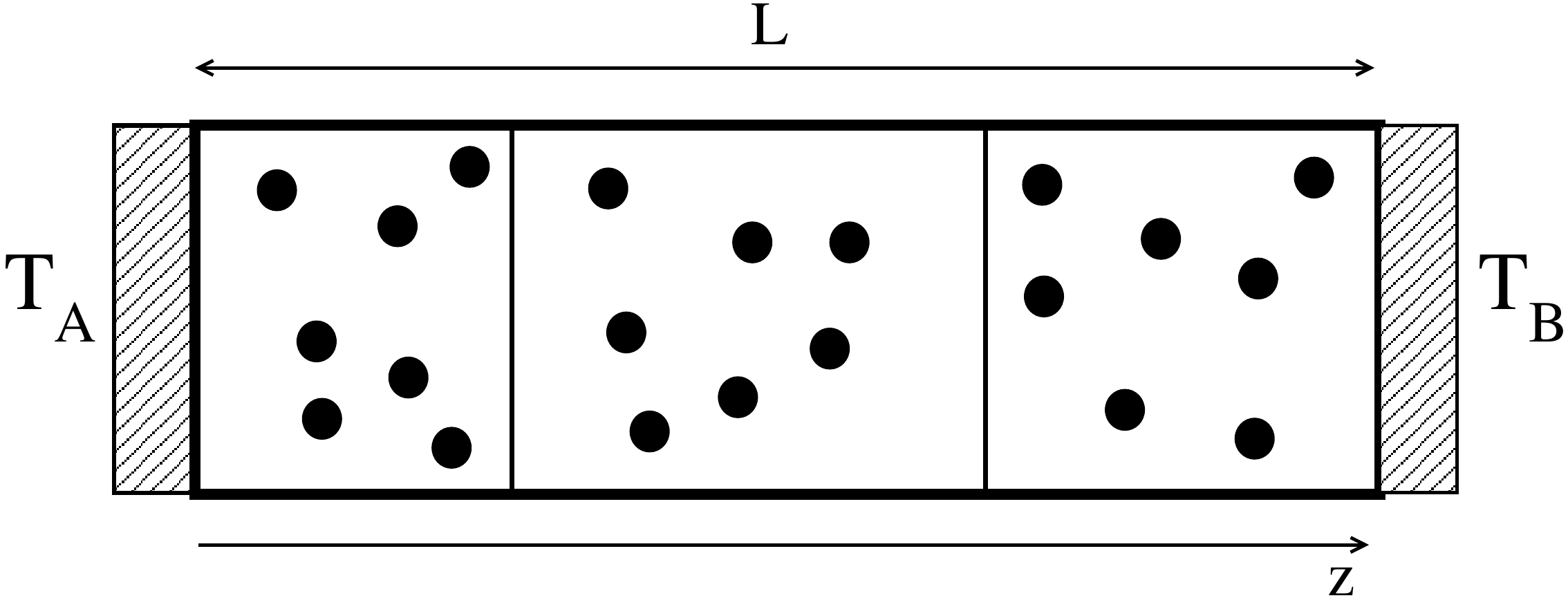}
\caption{Sketch of the system.}
\label{fig0}
\end{figure}

We consider a box of length $L$, partitioned by $n$ mobile
adiabatic walls (also called pistons in the following), with mass $M$,
average positions $z_j$ and velocities $V_j$ ($j=1, 2, \dots n$).  The walls
move without friction along the horizontal axis (see Figure~\ref{fig0}). The
external walls are kept fixed in the positions $z_0=0$ and
$z_{n+1}=L$.  Each of the $n+1$ compartments separated by the walls
contains $N$ non-interacting point-like particles, with mass $m$,
positions $x_i$ and velocities $v_i$.  The particles interact with the
pistons via elastic collisions:
\begin{equation}
\label{eq:urti_elastici}
\begin{aligned}
V' &= V + \frac{2\,m}{m+M} \, (v_i - V), \\
v'_i &= v_i - \frac{2\,M}{m+M} \, (v_i - V),
\end{aligned}
\end{equation}
where primes denote post-collisional velocities.

The two external walls in $z=0$ and $z=L$ act as thermal baths at temperatures
$T_A$ and $T_B$.  The interaction of the thermostats with the particles is
the following: when a particle collides with the wall, it is
reinjected into the system with a new velocity drawn from the
probability distribution~\cite{TTKB1998}
\begin{equation}
\label{eq:distribuzione_particelle_termostato}
\rho_{A, B}(v') = \frac{m}{k_B T_{A, B}} |v'| \,e^{-\frac{mv'^2}{2k_B T_{A, B}}} \,\Theta(\pm v'),
\end{equation}
with $+$ for the case $A$ and $-$ for $B$, and $k_B$ is the Boltzmann constant, $\Theta(x)$ being the Heaviside step~function:
\begin{equation}
\Theta(x)= 
\begin{cases}
1, & \quad\text{if} \quad x>0,\\
0, & \quad\text{if} \quad x\leq 0.
\end{cases}
\end{equation}

For the following discussion, we define the temperature of the $i-$th piston as
\begin{equation}
T_i^{(p)}=M\langle V_i^2\rangle,
\end{equation}
and the temperature of the particle gas in the compartment $j$
as an average on the particles between the $(j-1)$-th and the $j$-th piston 
\begin{equation}
T_j=m \frac{1}{N}\sum_{i\in (j-1)N}^{jN}\langle v_i^2\rangle.
\end{equation}

\section{Simple Thermodynamic Considerations}
\label{two}

We expect, and this is fully in agreement with the numerical
computations, that, given a generic initial condition, after a certain
transient, the system reaches a stationary state.  The positions of the
walls fluctuate around their mean values $z_j$, in a similar way to
$T_i^{(p)}$ and $T_i$.  The first non-trivial problem of the non
equilibrium statistical mechanics is to determine $z_j$, $T_i^{(p)}$
and $T_i$ as function of the parameters of the model, i.e., $n, L, m,
M, T_A$ and $T_B$.

We first analyze the simplest case of a single piston, where
thermodynamics is sufficient to provide a complete description of the
stationary state. Then, we consider the more general case of a multiple
piston; in such a situation, thermodynamic relations are not enough to
univocally determine the steady state: it is necessary to adopt a
statistical mechanics approach. Such an approach will
  rely on three main assumptions, discussed in more detail below:
  small mass ratio $m/M\ll 1$, local thermodynamic equilibrium in each
  compartment, and independence of the collisions particles/pistons.

\subsection{Single Piston}

If $n=1$, using the equation for the perfect gas in each compartment,
we immediately get the~equations
\begin{equation}
\label{eq:gas_perfetti}
\begin{cases}
p z_1  = N k_B T_A,\\
p (L-z_1) = N k_B T_B, 
\end{cases}
\end{equation}
where $p$ is the pressure, yielding
\begin{equation}
\label{eq:z_1(TA, TB)}
z_1 = \frac{T_A}{T_A+T_B}L.
\end{equation}
Therefore, in this case, thermodynamics univocally determines the
stationary state of the system.

\subsection{Multiple Piston}

We now consider the generalized piston model with $n>1$. An analysis
of the case $n=2$ is enough to understand how the relations obtained
from thermodynamics can be not sufficient to fully characterize the
non equilibrium steady state.  Indeed, we have the following relations:
\begin{equation}\label{eq:gas_perfetti_2}
\begin{cases}
p z_1  = N k_B  T_A,\\
p(z_2 - z_1) = N k_B T_1,\\
p(L-z_2) = N k_B T_B,
\end{cases}
\end{equation}
which give the constraints
\begin{equation}
\label{eq:z1z2(TATBT1)}
\begin{aligned}
z_1 = \frac{T_A}{T_A + T_1 + T_B}L,\\
z_2 = \frac{T_A+T_1}{T_A + T_1 + T_B}L.
\end{aligned}
\end{equation}
Therefore, since we have three variables ($z_1,z_2,T_1$) and only two
constraints, thermodynamics is not enough to determine the stationary
state. The computation can be easily extended to an arbitrary value of
$n$, leading to
\begin{equation}
\begin{cases}
pz_1 = N k_B T_A,\\
p(z_2 -z_1) = N k_B T_1\\
...,\\
p(z_n - z_{n-1}) = N k_B T_{n-1},\\
p(L-z_n) = N k_B T_B,
\end{cases}
\end{equation}
that give the conditions
\begin{myequation1}
\label{eq:posizioni_previsione_termodinamica}
\begin{aligned}
&z_1 = \frac{T_A}{T_A+\sum_{k=1}^{n-1}T_k+T_B }L, \quad \cdots \quad z_m = \frac{T_A+\sum_{k=1}^{m-1}T_k}{T_A+\sum_{k=1}^{n-1}T_k+T_B} L, \quad \cdots \quad
z_n = \frac{T_A+\sum_{k=1}^{n-1}T_k}{T_A+\sum_{k=1}^{n-1}T_k+T_B}L,
\end{aligned}
\end{myequation1}
with $m=2,\ldots, n-1$, namely, only $n$ constraints for $2n-1$ variables.

\section{Coarse-Grained Description and Effective Langevin Equations}
\label{three}

In order to obtain a statistical description of the system, we now
derive effective stochastic equations, governing the dynamics of the
relevant variables.  Previous theoretical studies based on a Boltzmann
equation approach on similar systems were reported
in~\cite{GP99,GPL03,GL05}. In particular, a generalized piston model
was considered in Reference~\cite{GL05}. In those works, theoretical results
were not compared to numerical simulations, so that the underlying
hypotheses and their range of validity remained~unclear.

Here, we present a different analysis.  We will assume that the
evolution of the observables is described by Langevin equations. The
idea, coming back to Smoluchowski, is to integrate out the fast
degrees of freedom of the gas particles by computing conditional
averages, knowing the macroscopic variables: position $z$ and velocity
$V$ of each piston, and temperatures $T$ of the gases. In order to
simplify the notation, let us denote by $\left<\cdot \right>$ this
average, meaning the conditional average $\left<\cdot |z, V,
T\right>$. We will compute the average change of a generic observable
$X$ in a small time interval $\Delta t$, due to the collisions between
the particles of the gas and the pistons.

Let us assume that in the stationary state the particles of the gas, in
each compartment, have uniform space distribution $\rho$ and a
Maxwell--Boltzmann distribution $\phi(v)$ at temperature $T$:
\begin{equation}
\label{eq:maxwell-Boltzmann}
\phi(v)=\MB,\qquad \rho(x) = \frac{1}{\Delta z},
\end{equation}
where $\Delta z$ is the length of the box containing the gas.  We can
obtain the rate of the collisions by considering the following
equivalent problem: piston at rest and a particle, which moves with the
relative velocity $v-V$. The point particles which collide against the
piston in $x$ in the time interval $dt$ are:
\begin{equation}
\label{eq:urto_dx_sx}
N\rho(x)\URTODX, \quad\quad N\rho(x)\URTOSX,
\end{equation}
respectively, for particles on the left and on the right with respect
to the piston. The Heaviside function $\Theta$ is necessary to have a
collision.

Let us now consider a generic observable $X_j$, depending in general on the
velocities of the gas particles and of the pistons. We want to write down a Langevin equation:
\begin{equation}
\frac{d X_j}{dt}=D_j({\bf X}) + \textrm{noise},
\end{equation}
where both the drift term $D_j({\bf X})$ (${\bf X}$ being the vector of all relevant macroscopic variables in the system)
and the noise term are due to 
collisions with the particles of the gas. We have:
\begin{equation}\label{eq:general_lr}
D_j({\bf X}) =\lim_{\Delta t \rightarrow 0} \frac{\left<\Delta X_j|{\bf X}\right>}{\Delta t} =\lim_{\Delta t \rightarrow 0} \frac{1}{\Delta t}\left[\left<\Delta X_j\right>^L_{coll} + \left<\Delta X_j\right>^R_{coll}\right],
\end{equation}
where
\begin{equation}
\begin{aligned}
\label{eq:general_left}
\left<\Delta X_j\right>^R_{coll}&=N\intpiu dv \, \int_{\zL - (v-V)\Delta t}^{\zL} dx \, X_j\,\rho(x) \phi(v)\Theta(v-V) = \frac{N}{\Delta \zL} \intpiu dv \, X_j\phi(v)(v-V)\Delta t,
\end{aligned}
\end{equation}
\begin{myequation2}
\begin{aligned}
\label{eq:general_right}
\left<\Delta X_j\right>^L_{coll}&=N\intmeno dv \, \int_{\zR}^{\zR - (v-V)\Delta t} dx \, X_j\,\rho(x) \phi(v) \Theta(V-v) = \frac{N}{\Delta \zR}\intmeno dv \, X_j\,\phi(v)(V-v)\Delta t.
\end{aligned}
\end{myequation2}
$\left<\Delta X_j\right>^R_{coll}$ is the variation of the observable due to the collisions
with the particles that have the pistons on the left, while $\left<\Delta X_j\right>^L_{coll}$ denotes the variation due to the collisions with the particles
which have the piston on the right.


In the stationary state, the macroscopic variables $T$, $V$ and $z$ do not
depend on time. It means that the time derivative of the conditional
average of a generic observable $X$ is zero, namely:
\begin{equation}
\label{eq:stato_stazionario}
\lim_{\Delta t \rightarrow 0} \frac{1}{\Delta t}\left[\left<\Delta X_j\right>^L_{coll} + \left<\Delta X_j\right>^R_{coll}\right]= 0.
\end{equation}

By using a perturbation development in the small parameter
$\epsilon=m/M$, it is possible to derive the relations between the
average positions and the temperatures of the pistons and the
temperatures of the gas. In order to obtain this result, we need to
average on the piston velocity, which is a stochastic variable. Of
course, in the steady state, the odd momenta have
to be zero:
\begin{equation}
\left<  V^{2\alpha+1} \right>=0,\qquad \alpha=0,1,2,... 
\end{equation}
The idea is to compute the average force produced by the particles of
the gas, which collide against the piston, by computing the average
momentum exchanged in the collisions. 

The evolution of the velocity of the piston is governed by the
following stochastic differential~equation:
\begin{equation}
M \der V = F_{coll}(\Delta \zL, \Delta \zR, \TL, \TR, V) + K \eta,
\label{vel}
\end{equation}
where $\eta$ is zero-average white noise, usually $K$ is a constant, and
$F_{coll}$ is the average force which acts on the piston. This force
is due to the collisions with the gas on the left and on the right,
and depends on the average size of the box on the right and on the
left of the piston, $\Delta \zL$ and $\Delta \zR$, respectively, and
on the temperatures of the gas on the left and on the right, $\TL$ and
$\TR$. Now let us set $\Delta X=V'-V$ in Equation~\eqref{eq:general_lr}. As detailed in the Appendix, computing
the left and right contributions to $F_{coll}$ and expanding in the
small mass ratio $\epsilon=\sqrt{m/M}$, we obtain at the lowest
orders:
\begin{flalign}
\label{eq:V_zero}
&O(1):\quad Nk_B\left( \frac{\TL}{\Delta \zL} - \frac{\TR}{\Delta \zR}\right),&\\
\label{eq:V_uno0}
&O(\epsilon): \quad -2N\sqrt{2k_B} \sqrt{\frac{M}{\pi}} \left[ \frac{T_L^{1/2}}{\Delta \zL} + \frac{T_R^{1/2}}{\Delta \zR}  \right] V.&
\end{flalign}
In the steady state, when the time derivative vanishes, we have from
the order $O(1)$ the following relation between the temperatures of
the gas and the average length of the boxes:
\begin{equation}
\Delta \zR=\Delta \zL \frac{T_{R}}{T_{L}}.
\end{equation}
This relation is nothing but the one that can be derived from thermodynamics. 
From Equation~(\ref{eq:V_uno0}), we obtain the obvious result $\langle V\rangle=0$.

Analogous computations, reported in the Appendix, can be carried out
for the variance of the pistons and the gas temperatures.  In
particular, for the variable $\Delta X=M(V'^2-V^2)$, at lowest orders
in $\epsilon$, we obtain the relations:
\begin{flalign}
\label{eq:Vq_zero0}
&O(1): \quad N2k_B \left(\frac{T_L}{\Delta \zL} - \frac{T_R}{\Delta \zR} \right),&\\
\label{eq:Vq_uno0}
&O(\epsilon): \quad \frac{N}{\sqrt{\pi}}\frac{4\sqrt{2}}{M^{1/2}}\left[\left(\frac{(k_BT_L)^{3/2}}{\Delta \zL} + \frac{(k_BT)^{3/2}}{\Delta \zR}  \right)-MV^2\left(\frac{(k_BT_R)^{1/2}}{\Delta \zR}+\frac{(k_BT_L)^{1/2}}{\Delta \zL}\right)\right].&
\end{flalign}

In the steady state, by using the thermodynamic relations, the order
$O(1)$ Equation~\eqref{eq:Vq_zero0} is identically zero. After integrating
over all values of the piston velocity and by requiring that the order
$O(\epsilon)$ vanishes, we obtain a relation between the temperature
of the piston $T^{(p)}$ and those of the gases on the left and on the right:
\begin{equation}
T^{(p)}= M\left<V^2\right> = \left(\TR \TL\right)^{1/2}.
\label{Tp}
\end{equation}

Finally, a condition on the temperature of the gas as a function of
the velocity variance of the pistons on its left and on its right can
be obtained by considering the variable $\Delta X=m(v'^2-v^2)/N$ in
Equation~\eqref{eq:general_lr}. This gives (see Appendix) the relations:
\begin{flalign}
\label{eq:T_zero0}
&O(1): \quad \frac{2k_BT}{\Delta z} \left(V_L -V_R\right),\\
\label{eq:T_uno0}
&O(\epsilon): \quad \frac{\sqrt{2k_B}}{\Delta z}\frac{4}{\sqrt{\pi}}\frac{T^{1/2}}{M^{1/2}} \left[MV^2_R+MV^2_L - 2T\right].
\end{flalign}
By integrating over the velocity of the piston, because
$\left<V\right>$ is zero, the order $O(1)$ is identically zero. By
requiring that, in the stationary state, the order $O(\epsilon)$
vanishes, we eventually obtain:
\begin{equation}\
\begin{aligned}
&0 = \frac{\sqrt{2k_B}}{\Delta z}\frac{4}{\sqrt{\pi}}\frac{T^{1/2}}{M^{1/2}} \left[M\left<V^2_R\right>_0+M\left<V^2_L\right>_0 - 2T\right] \qquad \Longrightarrow \qquad T= \frac{M}{2} \biggl(\left<V^2_R\right> + \left<V^2_L\right> \biggr).
\end{aligned}
\end{equation}

As expected from thermodynamics, one can easily show that the gases in
contact with the thermostats reach the temperatures of the thermal
baths (see Appendix for a detailed explanation).

\section{Numerical Simulations}
\label{four}

In order to check the range of validity of the above approach, we have
performed extensive molecular dynamics simulations (using
an event driven algorithm) of the system, varying the relevant
parameters of the model and comparing the analytical prediction of
Section 3 with the actual numerical results. The system is initialized
with the pistons placed at equidistant positions, with zero velocity,
while the gas particles are randomly distributed in each box, with
velocities drawn from a Gaussian distribution at temperature
intermediate with respect to those of the thermostats. We have, however,
checked that the stationary state reached by the system is independent
of the initial conditions. All data presented in the following are
measured in the steady state, after the transient relaxation dynamics
from the initial state.

\begin{figure}[ht!]
\centering
\includegraphics[width=.5\columnwidth,clip=true]{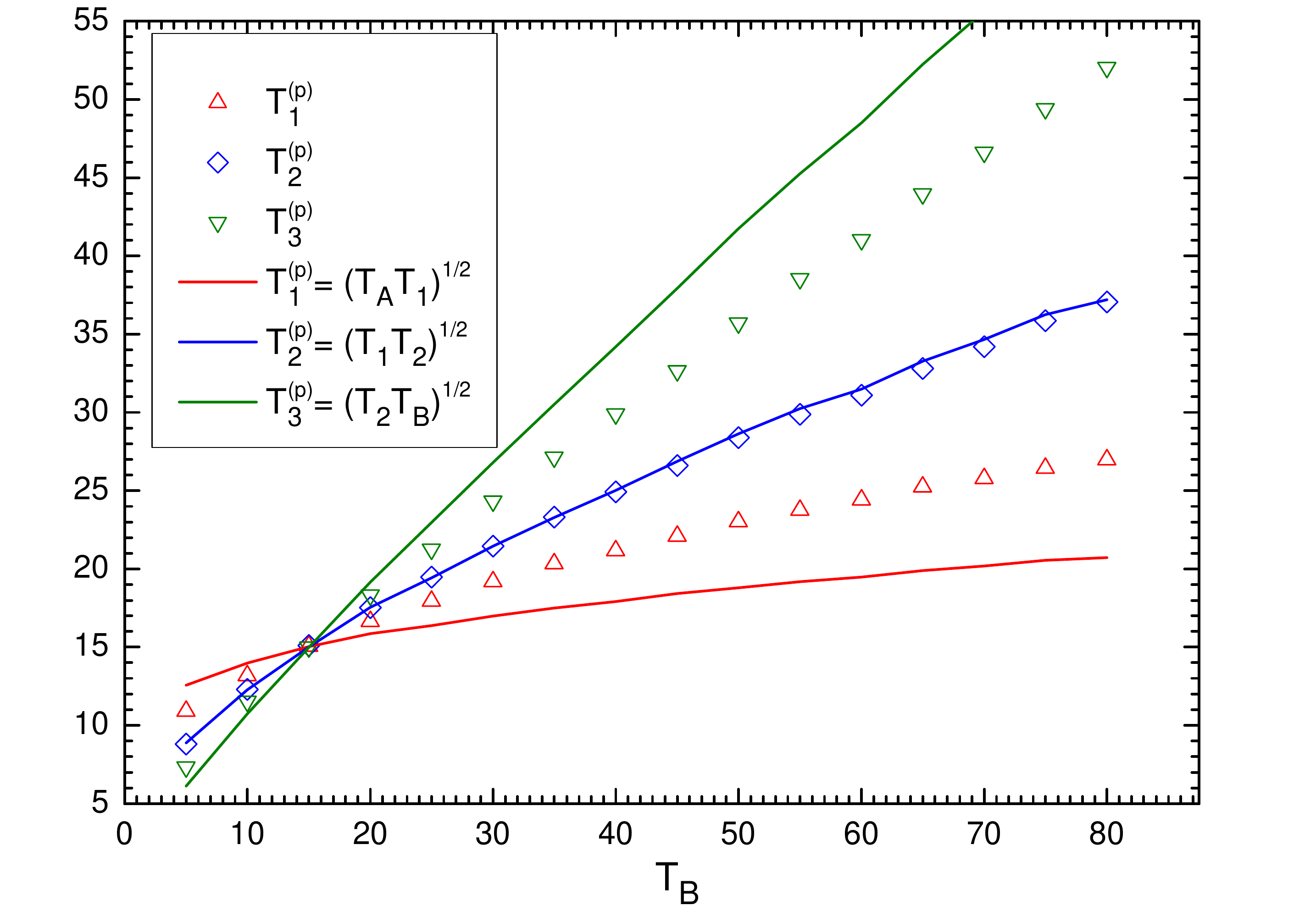}
\caption{Piston temperatures for a 3-piston system, with parameters $T_A=15$, \mbox{$N = M = 100$}, \mbox{$m = 1$}, \mbox{$L = 1$}.}
\label{fig1}
\end{figure}   

In Figure~\ref{fig1}, we show the piston temperatures $T_i^{(p)}$,
$i=1,2,3$, for a 3-piston system, as measured in numerical simulations
(symbols) for a certain choice of the parameters, and compare them
with the theoretical predictions of Equation~(\ref{Tp}) (lines). The
approximation is good for the intermediate piston, while it is not
very accurate for the side pistons, and is the worst in the case of a
large gradient $T_B/T_A\gg 1$. Simulations performed for systems with
more pistons give similar results.

\begin{figure}[ht!]
\centering
\includegraphics[width=.5\columnwidth,clip=true]{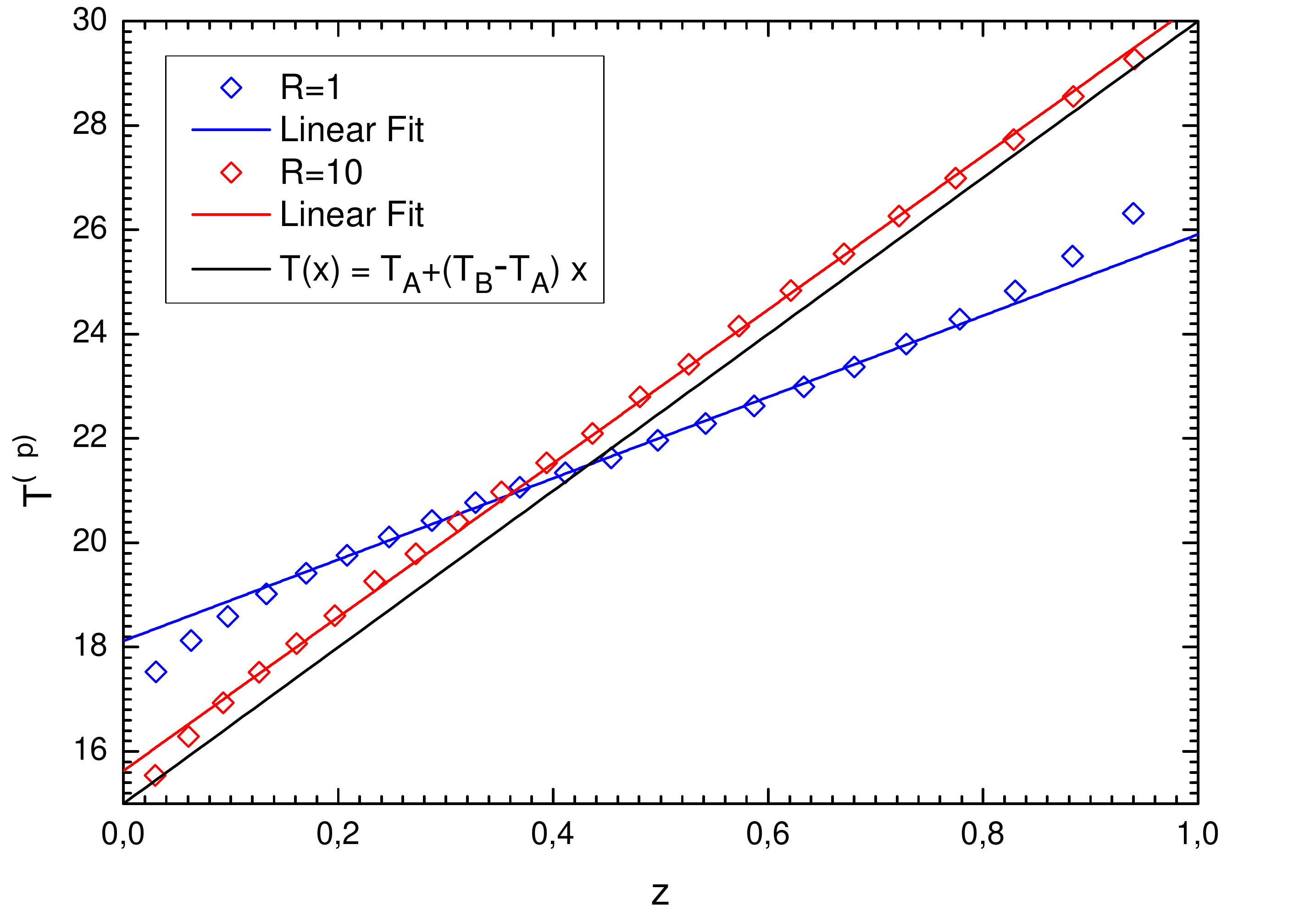}
\caption{Temperature profiles for a 22-piston system with $R=1$ and $R=10$. Other parameters are $T_A=15,T_B=30,M=50,m=1,L=1$.}
\label{fig2}
\end{figure}

In Figure~\ref{fig2}, we consider a system with a large number of pistons
(\emph{n} = 22), where the temperature profile shows a linear behavior, in
agreement with Fourier's law.  We report two sets of data, differing
in the value of the parameter $R=mN/M$. Notice that the linear
behavior extends for a wider range of $z$ in the case $R=10$. Lines
represent linear fits of the data. Let us note that, for large $n$ and
$R\gg 1$, $T^{(p)}$ vs. $z$ is in good agreement with a linear behavior
in the whole space interval (even close to the thermal~baths).

The derivation of the Langevin equation (see Appendix) is mainly based on the assumptions:
\begin{itemize}
\item $m/M\ll 1$;
\item a local thermodynamic equilibrium, i.e. in each compartment
  between the piston $i-1$ and the piston $i$, the spatial
  distribution of the particles is uniform and the velocity
  probability distribution is a Gaussian function, whose variance is
  given by the gas temperature;
\item the collisions particles/pistons are independent.
\end{itemize}

\begin{figure}[ht!]
\centering
\includegraphics[width=.5\columnwidth,clip=true]{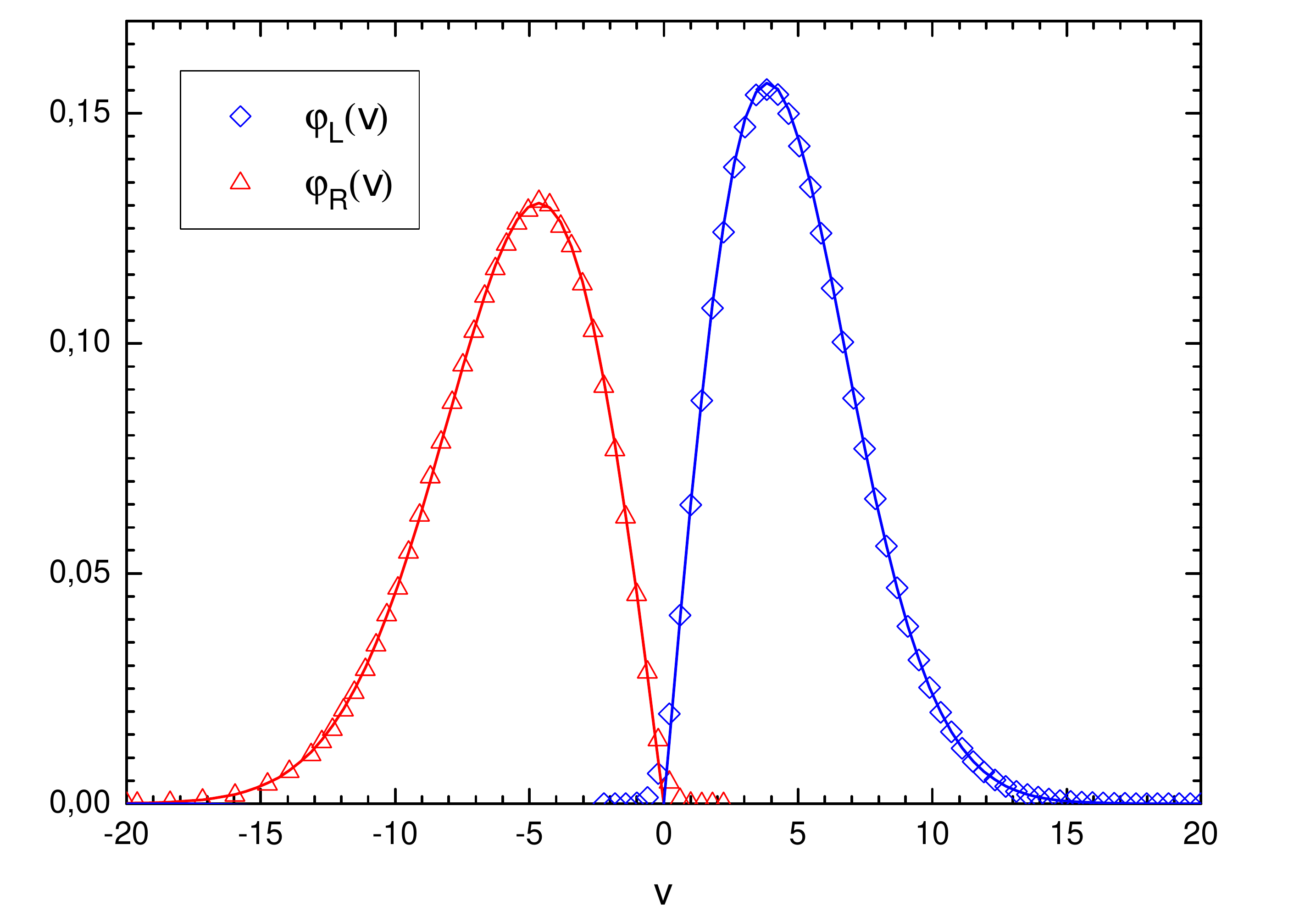}
\caption{Probability distribution function of the velocities
  $\phi_L(v)$ ($\phi_R(v)$) of the particles colliding with the first
  piston from the left (right), for a system with $n=2$. The curves,
  red and blue, are respectively: $\phi_L = \frac{m}{\KB \TL}
  e^{-v^2/\KB \TL}v\,\theta(-v)$ and $\phi_R = \frac{m}{\KB \TR}
  e^{-v^2/\KB \TR}v\,\theta(v)$, where $\TL=15.0$ and $\TR=21.6$ are
  the temperatures of the gas on the left and on the right with
  respect to the first piston. Other parameters are: $\TTL=15$, $\TTR=30$,
  $N=100$, $M=100$, $m=1$, $n=2$, $L=1$.}\label{fig4}
\end{figure}

The first assumption is easily checked, while the other ones are more
difficult. We can expect that a necessary condition for their validity
is that $N$ must be large. We have measured the velocity distribution
of the gas particles in the numerical simulations, finding that the
above hypothesis is verified. Moreover, we expect that
  the number of recollisions decreases with the ratio $m/M$, and we
  have numerically checked that their contribution is negligible (the
  fraction of recollisions on the total number of collisions is about
  $0.1\%$). Figure~\ref{fig4} shows the probability distribution
  functions of particles colliding from left, $\phi_L(v)$, and from
  right, $\phi_R(v)$, with a moving wall: the agreement with a
  Gaussian assumption is rather evident.

In addition, the numerical computations show that the left- and
right-moving particles in the same compartment have the following probability
distributions:
\begin{eqnarray}
\phi_-(v)&=&\sqrt{\frac{2m}{\pi k_B T_-}}\;\Theta(v)e^{-\frac{mv^2}{2k_BT_-}}, \\
\phi_+(v)&=&\sqrt{\frac{2m}{\pi k_B T_+}}\;\Theta(-v)e^{-\frac{mv^2}{2k_BT_+}}, \\
\end{eqnarray}
where $T_-$ ($T_+$) denotes the temperature of the particles incident
on the piston $i$ ($i-1$) from the left (right), with $T_-\approx T_+$, resulting
in a small but finite heat flow proportional to $(T_-- T_+)$, in agreement
with~\cite{GL05}. Let us note that the above probability distributions are different
from the distributions  $\phi_L(v)$ and  $\phi_R(v)$, which are computed for particles
actually colliding with the piston; this is the origin of the presence of the factor $v$
appearing in the expressions in the caption of Figure~\ref{fig4}.

The distribution of $v$ in the compartment between pistons $i-1$ and $i$, has the form:
$$
\phi(v) \propto A \Theta(v)e^{- \frac{mv^2}{ 2k_BT_-}} + B \Theta(-v) e^{ - \frac{mv^2}
{2k_BT_+}},
$$
where $A$ and $B$ are constants.
However, since
$$
T_- = T_+ + O(\epsilon^2),
$$
we have a Gaussian distribution for the particle gas in the compartment:
 \begin{equation}
\phi_i(v) \propto e^{- \frac{mv^2}{2k_BT_i}}+O(\epsilon^2), \qquad T_i = T_+ + O(\epsilon^2).
\end{equation}
Let us note that all our results are based on an expansion in powers of
$\epsilon$, neglecting $O(\epsilon^2)$. In other words, because the
heat rate is very small, this heat flux perturbs the Gaussian form of the
probability distribution $\phi$ in a negligible way.

\begin{figure}[ht!]
\centering
\includegraphics[width=.5\columnwidth,clip=true]{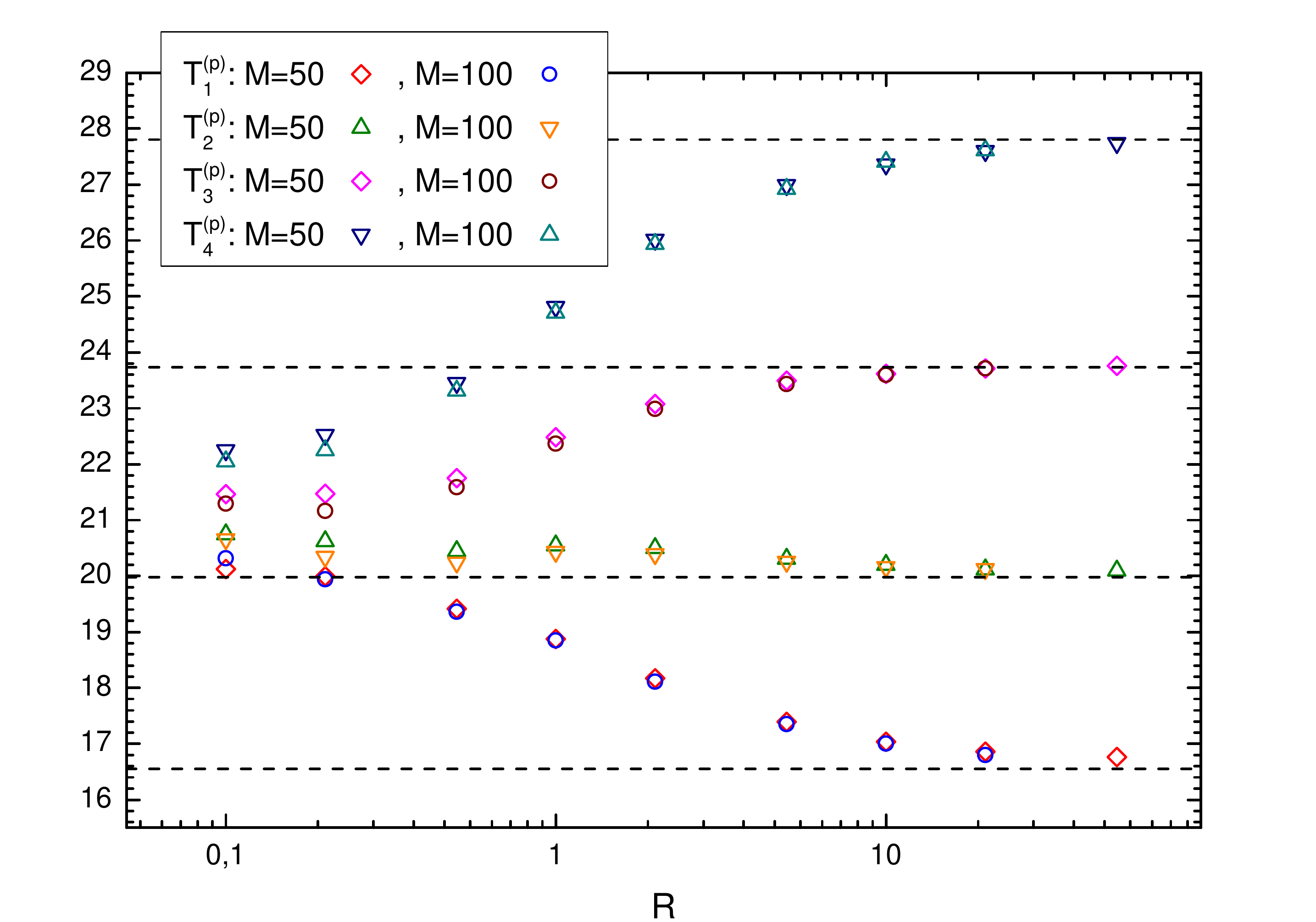}
\caption{Piston temperatures for a 4-piston system as a function of $R$. Dashed lines are the theoretical predictions. Other parameters are $T_A=15,T_B=30,m=1,L=1$.}
\label{fig3}
\end{figure}

In order to better understand the dependence of our theoretical
results on the parameter $R$, we have performed numerical simulations
of the system for different values of $R$. From the previous remark,
we expect to have an improvement of the agreement between the
numerical results and the analytical predictions by increasing $R$.
In Figure~\ref{fig3}, we compare the piston temperature in a 4-piston
system with theoretical predictions from Equation~(\ref{Tp}), finding a
very good agreement for large values of $R$. This shows that the total
mass of the gas contained in a box has to be larger than the mass of
the piston, in order for the kinetic theory presented in the previous
section to be accurate.

\section{Conclusions}

We have studied a generalized piston in contact with two thermal baths
at different temperatures.  This system represents a simple but
interesting case where the emergence of Fourier's law from a
microscopic mechanical model can be studied. We have presented a
kinetic theory treatment inspired by an approach \`a la Smoluchowski,
and we have investigated the range of validity of these results with
molecular dynamics numerical simulations. We have found that, in order
for the theory to be accurate, the ratio $R=mN/M$ should be large
enough, namely the total mass of the gas in each compartment should be
greater than that of the single piston.

We have considered ideal gas in our model, but we do not expect that
the introduction of short-range interactions among gas particles, at
least in not too dense cases, leads to significant differences in the
behavior of the system. An interesting, non-trivial, future line of
research in this model is the study of the relaxation  to the
steady state and the dynamical properties of the system, focusing on
correlation and response functions.

\appendix
\section{}
In this appendix, we will derive the relations that determine the
stationary state of the multi-component piston model described in the
main text. 

\subsection{Piston Position}

The effective force acting on the piston due to the collisions with gas particles, appearing in Equation~(\ref{vel}),
is given by two contributions, $F_{coll}=F^{L}_{coll}+F^{R}_{coll}$.
By taking into account the elastic collisions rule, Equation~\eqref{eq:urti_elastici}, these terms can be computed as follows:
\begin{myequation3}
\label{eq:V_left}
\begin{aligned}
F^{L}_{coll} &=\frac {N}{\Delta \zL} \intpiumeno dvM(v-V)\Theta(v-V) \phi(v) \,  (V' - V)= \frac{N}{\Delta \zL} \frac{2mM}{m+M} \intpiu dv \, (v - V)^2 \MBL \\
&= \frac{N}{\Delta \zL} \frac{M}{m+M}\left[\erfc \left(\sqrt{\normL} V\right)( mV^2+ k_B \TL) - \sqrt{\frac{m}{\pi}}\sqrt{2k_B\TL} V e^{-\normL V^2} \right],\\
\end{aligned}
\end{myequation3}
where in the first line we have used Equation~(\ref{eq:maxwell-Boltzmann}).
In the same way, by putting the observable $\Delta X=M(V'-V)$ in the Equation \eqref{eq:general_right}, we have:
\begin{myequation4}
\label{eq:V_right}
\begin{aligned}
F^{R}_{coll} &=\frac {N}{\Delta \zR} \intpiumeno dvM(V-v)\Theta(V-v) \phi(v) \,  (V' - V)= -\frac{N}{\Delta \zR} \frac{2mM}{m+M} \intmeno dv \, (v - V)^2 \MBR \\
&=- \frac{N}{\Delta \zR} \frac{M}{m+M}\Biggl[\left(1+\erf \left(\sqrt{\normR} V\right)\right)( mV^2+ k_B \TR) + \sqrt{\frac{m}{\pi}}\sqrt{2k_B\TR} V e^{-\normR V^2} \Biggr].
\end{aligned}
\end{myequation4}

By putting together the previous relations, we obtain:
\begin{myequation5}
\label{eq:V}
\begin{aligned}
M\der \left<V\right>_{coll} &= F^{L}_{coll} + F^{R}_{coll}= \frac{NM}{m+M}\left[\frac{k_B \TL}{\Delta z_L} \erfc \left (\sqrt{\normL} V\right) - \frac{k_B \TR}{\Delta \zR} \left (\erf \left(\sqrt{\normR} V\right)+1 \right) \right] -\\
&\,\,\,\, -\frac{NM}{m+M} \frac{\sqrt{2m}}{\pi} V \left[\frac{\sqrt{k_B \TL}}{\Delta \zL} e^{-\normL V^2}  + \frac{\sqrt{k_B \TR}}{\Delta \zR} e^{-\normR V^2} \right]+\\
&\,\,\,\, + \frac{NM}{m+M} mV^2 \left[ \frac{1}{\Delta \zL} \erfc \left(\sqrt{\normL} V\right) - \frac{1}{\Delta \zR} \left(1+\erf\left(\sqrt{\normR} V\right)\right)\right].
\end{aligned}
\end{myequation5} 

By expanding in the small ratio $\epsilon=\sqrt{m/M}\ll1$, and assuming that $V/v\sim \epsilon$, we have:
\begin{flalign}
\label{eq:V_zero1}
&O(1):\quad Nk_B\left( \frac{\TL}{\Delta \zL} - \frac{\TR}{\Delta \zR}\right),&\\
\label{eq:V_uno}
&O(\epsilon): \quad -2N\sqrt{2k_B} \sqrt{\frac{M}{\pi}} \left[ \frac{T_L^{1/2}}{\Delta \zL} + \frac{T_R^{1/2}}{\Delta \zR}  \right] V.&
\end{flalign}

In the steady state, when the time derivative vanishes, we have from
the order $O(1)$ the following relation between the temperatures of
the gas and the average lengths of the boxes:
\begin{equation}
\label{eq:thermodynamic_relation_appendix}
\Delta \zR=\Delta \zL \frac{T_{R}}{T_{L}}.
\end{equation}

\subsection{Piston Fluctuations}

Let us now consider the observable $\Delta X=M(V'^2 -V^2)$. From
Equation~\eqref{eq:general_left}, using Equation~\eqref{eq:urti_elastici}
for the elastic collisions, we obtain:
\begin{equation}
\begin{aligned}
M\der \left<V^2\right>^L_{coll} &= \frac{N}{\Delta \zL}\intpiumeno dv \,(v-V)\Theta(v-V)M(V'^2 - V^2) \phi(v)\\
&= \frac{NM}{\Delta \zL}\intpiu dv \,(v-V)\left[\frac{(2m)^2}{(m+M)^2}(V-v)^2+\frac{4m}{m+M}V(V-v) \right] \phi(v).
\end{aligned}
\end{equation}
Using Equation~(\ref{eq:maxwell-Boltzmann}), we have:
\begin{equation}
\label{eq:Vq_left}
\begin{aligned}
&M\der \left<V^2\right>^L_{coll} = \frac{MN}{\Delta \zL \sqrt{\pi}} \frac{4m^2}{(m+M)^2} \left(\invnormL\right)^{3/2}-\\
&\,\,\,\,-\factVqL \invnormL \sqrt{\pi} \erfc\left(\sqrt{\normL} V\right)\left[\frac{m}{m+M}\frac{3}{4}-\frac{1}{4}\right]V+\\
&\,\,\,\,+\factVqL \sqrt{\invnormL} e^{-\normL V^2} \left[\frac{m}{m+M}\frac{1}{2}-\frac{1}{2}\right]V^2-\\
&\,\,\,\,-\factVqL \frac{\sqrt{\pi}}{2} \erfc \left(\sqrt{\normL} V\right)\left[\frac{m}{m+M} -1\right]V^3.
\end{aligned}
\end{equation}
In the same way, from Equation~\eqref{eq:general_right}, we obtain:
\begin{equation}
\begin{aligned}
\label{eq:Vq_right}
M\der \left<V^2\right>^R_{coll} &= \frac{N}{\Delta \zR}\intpiumeno dv \,(V-v)\Theta(V-v)M(V'^2 - V^2) \phi(v)\\
&= \frac{NM}{\Delta \zR}\intmeno dv \,(V-v)\left[\frac{(2m)^2}{(m+M)^2}(V-v)^2+\frac{4m}{m+M}V(V-v) \right] \phi(v)\\
&= \frac{MN}{\Delta \zR \sqrt{\pi}} \frac{4m^2}{(m+M)^2} \left(\invnormR\right)^{3/2}+\\
&\,\,\,\,+\factVqL \invnormR \sqrt{\pi} \left(1+\erf\left(\sqrt{\normR} V\right)\right)\left[\frac{m}{m+M}\frac{3}{4}-\frac{1}{4}\right]V+\\
&\,\,\,\,+\factVqR \sqrt{\invnormR} e^{-\normR V^2} \left[\frac{m}{m+M}\frac{1}{2}-\frac{1}{2}\right]V^2-\\
&\,\,\,\,+\factVqR \frac{\sqrt{\pi}}{2} \left(1+\erfc \left(\sqrt{\normR} V\right)\right)\left[\frac{m}{m+M} -1\right]V^3.
\end{aligned}
\end{equation}
Putting together the contributions from the relations~\eqref{eq:Vq_left} and~\eqref{eq:Vq_right}, we have
\begin{equation}
\begin{aligned}
&M\der \left<V^2\right>_{coll}= \frac{MN}{\sqrt{\pi}}\frac{4m^2}{(m+M)^2}\left(\frac{2k_B}{m}\right)^{3/2}\frac{1}{2} \left[\frac{\TL^{3/2}}{\Delta \zL} e^{-\normL V^2} + \frac{\TR^{3/2}}{\Delta \zR}e^{-\normR V^2}  \right] +\\
&+ \factVq \frac{2k_B}{m}\left[\frac{3}{4}\frac{m}{m+M}-\frac{1}{4}\right]\Biggl\{\frac{\TR}{\Delta \zR} \left(\erf\left(\sqrt{\normR} V\right)+1\right)-\\
& -\frac{\TL}{\Delta \zL} \erfc\left(\sqrt{\normL} V\right)\Biggr\} V +\factVq \frac{1}{2\sqrt{\pi}}\sqrt{\frac{2k_B}{m}}\left[\frac{m}{m+M} -1\right] \\
& \times \left\{\frac{\sqrt{\TL}}{\Delta \zL} e^{-\normL V^2} + \frac{\sqrt{\TR}}{\Delta \zR} e^{-\normR V^2} \right\} V^2 +\factVq\frac{1}{2} \left[\frac{m}{m+M}-1\right] \\
& \times \left\{ \frac{1}{\Delta \zR}\left(\erf\left(\sqrt{\normR} V\right)+1\right) - \frac{1}{\Delta \zL}\erfc\left(\sqrt{\normL} V\right)\right\} V^3.
\end{aligned}
\end{equation}
By using a Taylor expansion in the small parameter $\epsilon = \sqrt{\frac{m}{M}}\ll1$, we obtain:
\begin{flalign}
\label{eq:Vq_zero}
&O(1): \quad 2N \left(\frac{\KB T_L}{\Delta \zL} - \frac{\KB T_R}{\Delta \zR} \right)V,&\\
\label{eq:Vq_uno}
&O(\epsilon): \quad \frac{N}{M^{1/2}}\frac{4\sqrt{2}}{\sqrt{\pi}}\left[\left(\frac{(k_BT_L)^{3/2}}{\Delta \zL} + \frac{(k_BT_R)^{3/2}}{\Delta \zR}  \right)-MV^2\left(\frac{(k_BT_L)^{1/2}}{\Delta \zL}+\frac{(k_BT_R)^{1/2}}{\Delta \zR}\right)\right].& 
\end{flalign}
In the steady state, by using the thermodynamic relations, the order $O(1)$ Equation \eqref{eq:Vq_zero} is identically zero. After integrating over all values of the piston velocity and by requiring that the order $O(\epsilon)$ vanishes, we obtain a relation between the temperature of the right piston and that of the left one:
\begin{equation}
T^{(p)}\equiv M\left<V^2\right> = \KB\left(\TR \TL\right)^{1/2},
\end{equation}
where we have used also the thermodynamic relation Equation~\eqref{eq:thermodynamic_relation_appendix}.

\subsection{Temperature of the Gas}

Let us now compute the average temperature of the gas. We have to distinguish between two different cases:
 
\begin{itemize}
\item Gas between two moving walls
\item Gas between a moving wall and a thermostat
\end{itemize}

\subsubsection{Gas between Two Moving Walls}

Let us consider the observable $\Delta X$ equal to the difference of the gas energy before and after the collision: $\Delta X=m(v'^2 - v^2)/N$.
By putting the observable into the relation \eqref{eq:general_left} and by taking into account the Equation \eqref{eq:urti_elastici}, we obtain:
\begin{equation}
\begin{aligned}
\frac{m}{N} \der \left<v^2\right>^L_{coll} &= \frac{m}{\Delta z}\intpiumeno dv\,(v-V_R)\Theta(v-V_R) [v'^2 - v^2] \phi(v)\\
&=\frac{m}{\Delta z} \intpiuR dv \, (v-V_R)\left[\frac{4M^2}{(m+M)^2}(V_R - v)^2 +\frac{4M}{m+M}v(V_R-v)\right]  \phi(v).
\end{aligned}
\end{equation}
By integrating, we have:
\begin{myequation6}
\label{eq:T_left}
\begin{aligned}
\frac{m}{N} \der \left<v^2\right>^L_{coll}&=\frac{m}{\Delta z} \frac{1}{\sqrt{\pi}}\frac{4M}{m+M} \left(\invnorm\right)^{3/2} e^{-\norm V_R^2}\left[-\frac{m}{m+M}\right]+\\
&+\frac{m}{\Delta z}\frac{4M}{m+M} \frac{2k_B T}{m}\erfc\left( \sqrt{\norm} V_R \right)\left[-\frac{M}{m+M}+\frac{2m}{m+M}\right]\frac{V_R}{4}+\\
&+\frac{m}{\Delta z}\frac{1}{\sqrt{\pi}}\frac{4M}{m+M} \sqrt{\frac{2k_BT}{m}}\frac{1}{2}\frac{M}{m+M} e^{-\norm V_R^2} V_R^2-\frac{m}{\Delta z} \frac{2M^2}{(M+m)^2}\erfc\left({\sqrt{\norm}V_R}\right) V_R^3.
\end{aligned}
\end{myequation6}
In the same way, we obtain:
\begin{equation}
\begin{aligned}
\label{eq:T_right}
\frac{m}{N} \der \left<v^2\right>^R_{coll} &= \frac{m}{\Delta z}\intpiumeno (V_L-v)\Theta(V_L-V) [v'^2 - v^2] \phi(v)\\
&=\frac{m}{\Delta z} \intpiuL dv \, (V_L-v)\left[\frac{4M^2}{(m+M)^2}(V_L-v)^2 +\frac{4M}{m+M}v(V_L-v)\right]  \phi(v)\\
&=\frac{m}{\Delta z} \frac{1}{\sqrt{\pi}}\frac{4M}{m+M} \left(\invnorm\right)^{3/2} e^{-\norm V_R^2}\left[-\frac{m}{m+M}\right]-\\
&-\frac{m}{\Delta z}\frac{4M}{m+M} \frac{2k_B T}{m}\left(1+\erf\left( \sqrt{\norm} V_L \right)\right)\left[-\frac{M}{m+M}+\frac{2m}{m+M}\right]\frac{V_L}{4}+\\
&+\frac{m}{\Delta z}\frac{1}{\sqrt{\pi}}\frac{4M}{m+M} \sqrt{\frac{2k_BT}{m}}\frac{1}{2}\frac{M}{m+M} e^{-\norm V_L^2} V_L^2+\\
&+2\frac{m}{\Delta z}  \frac{M^2}{(M+m)^2}\left(1+\erf\left({\sqrt{\norm}V_L}\right)\right) V_L^3.
\end{aligned}
\end{equation}
Putting together Equations~\eqref{eq:T_left} and~\eqref{eq:T_right}, one has:
\begin{equation}
\begin{aligned}
\frac{m}{N}\der \left< v^2\right>_{coll}&=-\frac{m}{\Delta z} \frac{1}{\sqrt{\pi}} \frac{4M}{m+M}\frac{m}{m+M} \left(\frac{2k_BT}{m}\right)^{3/2} \left[\frac{e^{-\norm V^2_L}}{2} +\frac{e^{-\norm V^2_R}}{2} \right]+\\
&+\frac{m}{\Delta z} \frac{4M}{m+M}\frac{2k_B T}{m} \frac{1}{4}\left[\frac{2m}{m+M} - \frac{M}{m+M}\right] \Biggl\{V_R \erfc\left(\sqrt{\norm} V_R\right) - \\ 
&-V_L - V_L \erf\left(\sqrt{\norm} V_L\right) \Biggr\} +\frac{m}{\Delta z} \frac{1}{2}\frac{4M^2}{(m+M)^2} \sqrt{\invnorm}\frac{1}{\sqrt{\pi}}  \times \\
&\times \left[V^2_R e^{-\norm V^2_R} + V^2_L e^{-\norm V^2_L} \right] -\frac{m}{\Delta z} \frac{4M^2}{(m+M)^2} \frac{1}{2} \times \\
&\times \left[V^3_R \erfc\left(\sqrt{\norm} V_R\right) - V^3_L \left(1+\erf\left(\sqrt{\norm} V_L\right)\right)\right],
\end{aligned}
\end{equation}
and using a Taylor expansion around the small parameter $\epsilon = \sqrt{\frac{m}{M}} \ll 1$ yields:
\begin{flalign}
\label{eq:T_zero}
&O(1): \quad \frac{2k_BT}{\Delta z} \left(V_L -V_R\right),\\
\label{eq:T_uno}
&O(\epsilon): \quad \frac{\sqrt{2k_B}}{\Delta z}\frac{4}{\sqrt{\pi}}\frac{T^{1/2}}{M^{1/2}} \left[MV^2_R+MV^2_L - 2T\right].
\end{flalign}
By integrating over the velocity of the piston, because
$\left<V\right>$ is zero, the order $O(1)$ is identically zero. By
requiring that in the stationary state the order $O(\epsilon)$
vanishes, we obtain a relation between the temperature of the gas and
those of the near pistons:
\begin{equation}\
\begin{aligned}
&0 = \frac{\sqrt{2k_B}}{\Delta z}\frac{4}{\sqrt{\pi}}\frac{T^{1/2}}{M^{1/2}} \left[M\left<V^2_R\right>_0+M\left<V^2_L\right>_0 - 2T\right] \qquad \Longrightarrow \qquad T= \frac{M}{2} \biggl(\left<V^2_R\right> + \left<V^2_L\right> \biggr).
\end{aligned}
\end{equation}

\subsubsection{Gas between a Piston and a Thermostat}
Consider the gas that is near a thermostat and piston. The
variation of the temperature due to the piston is the same as before
and is given by the Equations \eqref{eq:T_left} or \eqref{eq:T_right},
respectively, for a piston on the right or on left, with respect to the considered
gas. On the side of the thermostat, after the collision, the
particle takes a velocity according to the distribution given by the
Equation~\eqref{eq:distribuzione_particelle_termostato}.  For instance, if
the thermostat is that one on the left at temperature $T_0$, we have:
\begin{equation}
\begin{aligned}
\left<T\right>_{ther}^L &= \frac{m}{\Delta z} \intpiumeno dv \intpiumeno dv' \, (v'^2 -v^2)\phi(v)\,|v|\, \Theta(-v) \Phi_{T0}(v')\\
& \frac{m}{\Delta z} \intpiumeno dv \intpiumeno dv' \, (v'^2 -v^2)\MB\theta(-v)\,|v|\, v'\, \Theta(v') \frac{m}{k_B T_{0}}e^{-\frac{m}{2k_BT_0}v'^2} \\
& \frac{m}{\Delta z}\frac{m}{k_B T_{0}}\sqrt{\frac{m}{2\pi k_BT}}\int_{-\infinito}^0 dv \int_0^{\infinito}dv' v\,v'(v'^2 -v^2) e^{-\norm v^2} e^{-\frac{m}{2k_B T_0}v'^2}.
\end{aligned}
\end{equation}
By integrating, we obtain:
\begin{equation}
\label{eq:T_term}
\left<T\right>_{ther}^L=\sqrt{\frac{2}{\pi\,m}}\frac{k_B^{3/2}}{\Delta z} \sqrt{T}[T_0 - T].
\end{equation}
In order to compute the variation of the temperature, we have to put
together Equations~\eqref{eq:T_term} and~\eqref{eq:T_left}:
\begin{equation}
\begin{aligned}
\der \left<T\right> &= \left< T \right>_{ther}^L + \frac{m}{N} \der\left<v^2\right>_{coll}^{R} = \sqrt{\frac{2}{\pi\,m}}\frac{k_B^{3/2}}{\Delta z} \sqrt{T}[T_0 - T] + \\
&+\frac{m}{\Delta z} \frac{1}{\sqrt{\pi}}\frac{4M}{m+M} \left(\norm\right)^{3/2} \frac{e^{-\norm V_R^2}}{2}\left[-\frac{m}{m+M}\right]+\\
&+\frac{m}{\Delta z}\frac{4M}{m+M} \frac{2k_B T}{m}\erfc\left( \sqrt{\norm} V_R \right)\left[-\frac{M}{m+M}+\frac{2m}{m+M}\right]\frac{V_R}{4}+\\
&+\frac{m}{\Delta z}\frac{1}{\sqrt{\pi}}\frac{4M}{m+M} \sqrt{\frac{2k_BT}{m}}\frac{1}{2}\frac{M}{m+M} e^{-\norm V_R^2} V_R^2-\\
&-\frac{m}{\Delta z} 2 \frac{M^2}{(M+m)^2}\erfc\left({\sqrt{\norm}V_R}\right) V_R^3.
\end{aligned}
\end{equation}
By solving perturbatively around the small parameter $\epsilon = \sqrt{\frac{m}{M}}\ll 1,$ we obtain:
\begin{flalign}
\label{eq:T_menouno}
&O\left(\frac{1}{\epsilon}\right): \quad \sqrt{\frac{2}{\pi}} \frac{k_B^{3/2}}{M^{1/2}}\frac{\sqrt{T}}{\Delta z}(T_0 - T),&\\
&O(1): \quad -2\frac{\KB T}{\Delta z} V_R.&
\end{flalign}
In the steady state, from relation \eqref{eq:T_menouno} at order
$O(1/\epsilon)$, we have $T=T_0$.  This means that a gas near a
thermostat reaches the temperature of the thermal bath: this is the
result one can obtain from thermodynamics.  We can obtain exactly the
same result for a thermostat on the right with respect to the
considered gas.

\end{document}